\documentclass[twocolumn,showpacs,preprintnumbers,amsmath,amssymb]{revtex4}

 \def\theequation{\arabic{section}.\arabic{equation}}
\newcounter{rown}
\def\bl{\setcounter{rown}{\value{equation}}
        \stepcounter{rown}\setcounter{equation}0
        \def\theequation{\arabic{section}.\arabic{rown}\alph{equation}}}
\def\el{\setcounter{equation}{\value{rown}}
        \def\theequation{\arabic{section}.\arabic{equation}}}

\def\dalpha{{\dot\alpha}}
\def\dbeta{{\dot\beta}}

\def\dgamma{{\dot\gamma}}

\def\oomega{{\overline{\omega}}}
\def\oeta{{\overline{\eta}}}
\def\opi{{\overline{\pi}}}
\def\osigma{{\overline{\sigma}}}
\def\of{{\overline{f}}}
\def\orho{\overline{\rho}}

\def\weta{{\widehat{\eta}}}
\def\woeta{{\widehat{\oeta}}}
\def\wP{{\widehat{P}}}
\def\wsigma{{\widehat{\sigma}}}
\def\wosigma{{\widehat{\osigma}}}

\def\wrho{{\widehat{\rho}}}
\def\worho{{\widehat{\orho}}}
\def\wtt{{\widehat{\mathbf{t}^2}}}
\def\wt3{{\widehat{t}_3}}

\def\ada{{\alpha\dalpha}}
\def\adb{{\alpha\dbeta}}

\def\oZ{\overline{Z}}

\begin{document}

\title{Massive relativistic particle model with spin from free
two-twistor dynamics\\and its quantization\\ \ \\\small{\it
Dedicated to the memory of our collaborator Andreas Bette}}

\author{
Jos\'e A. de Azc\'arraga$^{\dagger}$, Andrzej Frydryszak$^{\ast}$,
Jerzy Lukierski$^{\ast}$ and C\`{e}sar
Miquel-Espanya$^{\dagger}$$^{\ast}$}

\bigskip

\address{$^{\dagger}$Departamento de
F\'{\i}sica Te\'orica, Univ.~de Valencia and IFIC (CSIC-UVEG),
46100-Burjassot (Valencia), Spain
\\
$^{\ast}$Institute of Theoretical
 Physics, Wroc{\l}aw University, 50-204 Wroc{\l}aw, Poland}

\def\theequation{\arabic{section}.\arabic{equation}}

\begin{abstract}
We consider a relativistic particle model in an enlarged
relativistic phase space, \sloppy $\mathcal{M}^{18}=
(X_\mu,P_\mu,\eta_\alpha,\oeta_\dalpha,\sigma_\alpha,\osigma_\dalpha,e,\phi)$,
which is derived  from the free two-twistor dynamics. The spin
sector variables
$(\eta_\alpha,\oeta_\dalpha,\sigma_\alpha,\osigma_\dalpha)$ satisfy
two second class constraints and account for the relativistic spin
structure, and the pair $(e,\phi)$ describes the electric charge
sector. After introducing the Liouville one-form on
$\mathcal{M}^{18}$, derived by a non-linear transformation of the
canonical Liouville one-form on the two-twistor space, we analyze
the dynamics described by the first and second class constraints. We
use a composite orthogonal basis in four-momentum space to obtain
the scalars defining the invariant spin projections. The
first-quantized theory provides a consistent set of wave equations,
determining the mass, spin, invariant spin projection and electric
charge of the relativistic particle.  The wavefunction provides a
generating functional for free, massive higher spin fields.
\end{abstract}


\maketitle

\section{Introduction}\label{Introduction}
\setcounter{footnote}{0} The choice of the basic geometric variables
that describe the most fundamental dynamics is an important and open
question. In particular, by analogy with the `hidden' quark
structure of hadronic particles, one may assume that the spacetime
coordinates are composite variables that can be expressed in terms
of more basic, fundamental geometric spinorial variables. This is
the twistor theory approach, proposed by Penrose in $D=4$ (see {\it
e.g.} \cite{Pen68}-\cite{Tod77}) which has  been  generalized to
higher dimensions (see {\it e.g.} \cite{Unknown-6,Unknown-7}) as
well as supersymmetrized (see {\it e.g.} \cite{Fer78,Luxx}).

Most of the studies of twistor dynamics have been restricted to
the one-twistor geometry, based on four complex spinorial
coordinates $Z_A=(Z_1,Z_2,Z_3,Z_4)$ supporting a fundamental
representation of the spinorial covering $SU(2,2)\simeq
\overline{SO(4,2)}$ of the $D=4$ conformal group. A twistor is
described by a pair of $D=4$ Lorentz  spinors,
\begin{eqnarray}
Z^A&=&(\omega^\alpha,\opi_\dbeta)\quad,\nonumber\\ \quad
\overline{Z}_A&=& (\pi_\alpha,\oomega^\dbeta)\ , \ A=1,2,3,4;\;
\alpha,\beta=1,2 \  .\label{OneTwistorDef}
\end{eqnarray}
In the one-twistor framework  the four-momentum is introduced as
follows
\begin{equation}
\label{OneTwistorPDef} P_{\alpha\dot\beta}=\pi_\alpha\opi_\dbeta
\qquad
(P_\mu=\frac{1}{\sqrt{2}}(\sigma_\mu)^{\alpha\dbeta}P_{\alpha\dbeta})
\quad.
\end{equation}
If we recall the Penrose basic relations \cite{Pen68,Pen72}
\begin{equation}
\label{OneTwistorPenroseRelations}
\omega^\alpha=iz^{\alpha\dbeta}\opi_\dbeta\qquad , \qquad
\oomega^\dalpha=-i\pi_\beta \overline{z}^{\beta\dalpha}\quad ,
\end{equation}
and assume that the $SU(2,2)$ twistor norm is zero,
\begin{equation}\label{OneTwistorNorm}
\left<Z,\overline{Z}\right>=
\omega^\alpha\pi_\alpha+\oomega^\dalpha\opi_\dalpha=0\quad ,
\end{equation}
the Minkowski space coordinates
$z^{\alpha\dbeta}=x^{\alpha\dbeta}+iy^{\alpha\dbeta}\equiv
\frac{1}{\sqrt{2}}(\sigma_\mu)^{\adb}z^\mu$ in formula
(\ref{OneTwistorPenroseRelations}) become real (\textit{i.e.}
$y^{\alpha\dbeta}=0$).
 Using (\ref{OneTwistorPDef}), (\ref{OneTwistorPenroseRelations})
 and $y^{\adb}=0$ one can show
the equivalence of the following three Liouville one-forms:
\begin{equation}\label{OneTwistorLiouvilleEquivalence}
\Theta^{(1)}=P_{\alpha\dbeta}dx^{\alpha\dbeta}=\pi_\alpha\opi_\dbeta
dx^{\alpha\dbeta}=\frac{i}{2}\left(\omega^\alpha d\pi_\alpha+\opi_\dalpha
d\oomega^\dalpha\right)+c.c.
\end{equation}

If we observe that the vanishing mass condition ($P^2=0$) follows
from (\ref{OneTwistorPDef}) and take into account eqs.
(\ref{OneTwistorDef}) and (\ref{OneTwistorLiouvilleEquivalence}),
 the following three models
describing a free massless scalar relativistic particle are seen to be equivalent:
\begin{widetext} \bl
\begin{eqnarray}
\mathcal{L}_1&=& \int d\tau \left(P_\mu \dot x^\mu-\lambda
P^2\right) \quad\text{(relat. phase space description)}\;;\qquad\label{OneTwistorRelPhaSpaPicture}\\
\mathcal{L}_2&=&\int d\tau\pi_\alpha\opi_\dbeta \dot x^{\alpha\dbeta}\quad\text{(mixed
space-twistor description \cite{Shi83})}\;;\qquad\label{OneTwistorMixedPicture}\\
\mathcal{L}_3&=&\int d\tau \left(\overline{Z}_A\dot
Z^{A}-\lambda\left<Z,\overline{Z}\right> \right)\quad \text{(pure
twistor description)}\ .\qquad\label{OneTwistorPurePicture}
\end{eqnarray}
\el
\end{widetext}

To describe particles with mass and spin in a twistor formalism,
it is necessary to use a pair of twistors \cite{Per77,Hu79,Tod77}
\begin{equation}\label{TwoTwistorDef}
Z^A_{\phantom{A};1}=(\omega^\alpha,\opi_\dbeta)\qquad , \qquad
Z^A_{\phantom{A};2}=(\lambda^\alpha,\oeta_\dbeta)\quad .
\end{equation}
In comparison with the one-twistor description of spacetime,
 the use of two twistors produces important changes:

i) the spacetime coordinates $x_\mu$ can be considered as
two-twistor composites;

ii) the appearance of non-vanishing spin and internal (electric)
charge leads to the complexification of the spacetime coordinates.

The composite complex Minkowski coordinates
$z_\mu=x_\mu+iy_\mu\quad
(z_\mu=\frac{1}{\sqrt{2}}(\sigma_\mu)_{\alpha\dbeta}z^{\alpha\dbeta})$
are described by the well known Penrose formula:
\begin{equation}\label{ComplexZFormula}
z^{\alpha\dbeta}=
\frac{i}{f}(\omega^\alpha\oeta^\dbeta-\lambda^\alpha\opi^\dbeta)\quad ,
\end{equation}
where
\begin{equation}\label{fDef}
f=\opi^\dalpha\oeta_\dalpha\quad .
\end{equation}

{}From the free two-twistor Liouville one-form
($\oZ_A^{\phantom{A};i}=(Z_{A;i})^*, i=1,2$)
\begin{equation}\label{TwoTwistorLiouvilleInitial}
\Theta^{(2)}=
\frac{i}{2}\left(Z^A_{\phantom{A};j}d\oZ_A^{\phantom{A};j}-\oZ^{A;j}dZ_{A;j}\right)
\quad,
\end{equation}
one obtains the following twistorial Poisson brackets (TPB) \bl
\begin{eqnarray}\label{PoissonBracketsFund1}
\left\{\pi_\alpha , \omega^\beta\right\} =& i
 \delta^{\beta}_{\alpha}\quad  , &\quad
 \left\{\eta_\alpha , \lambda^\beta\right\} =
 i \delta^{\beta}_{\alpha}\, ,\\
 \left\{\overline{\pi}_{\dot\alpha} ,
 \overline{\omega}^{\dot\beta}\right\} =& -i
 \delta^{\dot\beta}_{\dot\alpha}\quad , &
 \quad \left\{\overline{\eta}_{\dot\alpha} ,
 \overline{\lambda}^{\dot\beta}\right\} =
 -i \delta^{\dot\beta}_{\dot\alpha}\, .
 \label{PoissonBracketsFund2}
\end{eqnarray}
\el

Using (\ref{ComplexZFormula}) it follows that the spacetime
coordinates $x_\mu=\mathrm{Re}\, z_\mu$ have non-vanishing TPB's
\cite{Be84}
\begin{equation}\label{XNonCommutative}
\{x_\mu,x_\nu\}=-\frac{1}{(2|f|^2)^2}\epsilon_{\mu\nu\rho\tau}W^{\rho}P^{\tau}\quad,
\end{equation}
where $P_\mu$  is the two-twistor composite momentum,
\begin{equation}\label{TwoTwistorPDef}
P_{\alpha\dbeta}=\pi_\alpha\opi_\dbeta+\eta_\alpha\oeta_\dbeta\quad,
\end{equation}
\begin{equation}\label{P2}
P^2=P_{\adb}P^\adb=2(\pi_\alpha\eta^\alpha)(\opi_\dalpha\oeta^\dalpha)=2|f|^2\quad,
\end{equation}
and $W_\mu$ is the composite Pauli-Luba\'{n}ski spin four-vector,
$W^\mu P_\mu=0$. If we denote $\pi_\alpha\equiv \pi_{\alpha;1},\
\eta_\alpha\equiv\pi_{\alpha;2}$,
$\opi_\dalpha\equiv\opi_{\dalpha}^{\phantom{\alpha};1},\
\oeta_\dalpha\equiv\opi_\dalpha^{\phantom{\alpha};2}$ and
introduce
\begin{equation}\label{TwoTwistorPRDef}
P^{(r)}_{\alpha\dbeta}=
(\pi_{\alpha;i}\opi_{\dbeta}^{\phantom{\beta};j})(\tau^r)_j^{\phantom{j}i}
\qquad,\qquad r=1,2,3 \qquad,
\end{equation}
then $W_\mu$ in (\ref{XNonCommutative}) is given by the formula
\begin{equation}\label{RelationPLMomentum}
W_{\alpha\dbeta}=t^r P^{(r)}_{\alpha\dbeta}\quad,
\end{equation}
\begin{equation}\label{TijDef}
t^r=(\tau^r)_i^{\phantom{i}j}t_j^{\phantom{j}i} \ \quad
t_i^{\phantom{i}j}=Z^A_{\phantom{A};i}\oZ_A^{\phantom{A};j}\quad,
\end{equation}
where $\tau^r\equiv \sigma^r$ are the three $2\times 2$ Pauli
matrices satisfying the standard algebra
$(\tau^r)_i^{\phantom{i}k}(\tau^s)_k^{\phantom{k}j}=
\delta^{rs}\delta_i^j+i\epsilon^{rst}(\tau^t)_i^{\phantom{i}j}$.
The three composite four-vectors $P_\mu^{(r)}$ (eq.
(\ref{TwoTwistorPRDef})) together with $P^{(0)}_\mu\equiv P_\mu$
(eq. (\ref{TwoTwistorPDef})) describe an orthonormal basis in
four-momentum space ($\eta_{AB}=(+,-,-,-)$)
\begin{equation}\label{POrthNormBasis}
P^{(A)}_\mu P^{\mu\,(B)}=P^2\eta^{AB}\quad,\qquad A,B=0,1,2,3
\quad.
\end{equation}

In order to complete the set of observables we  add
 to the spin four-vector square $W_\mu W^\mu$
the projection of the relativistic spin vector $W_\mu$ on any
orthogonal direction to $P_\mu$. Taking into consideration the
orthogonal basis (\ref{POrthNormBasis}), the Lorentz invariant
spin projection
  can be represented by the scalar product of
$S_\mu=\frac{W_\mu}{\sqrt{P^2}}$ and the four-vector
$P_\mu^{(3)}$, normalized to unity. Denoting the projection by
$S^{(3)}$ and using (\ref{RelationPLMomentum}) one obtains
\begin{equation}\label{S(3)}
S^{(3)}=-\frac{1}{\sqrt{P^2}}P_\mu^{(3)}S^\mu=
       -\frac{1}{P^2}P^{(3)}_\mu W^\mu=t^3\quad.
\end{equation}

In general, we find
\begin{equation}\label{S-R}
S^{(r)}=-\frac{1}{P^2}P^{(r)}_\mu W^\mu=t^r \qquad, \qquad r=1,2,3
\qquad ,
\end{equation}
where $t^r$ denote the three scalars that provide the three
invariant spin projections.  Thus, contrarily to the standard
choice of the projection on a fixed space direction ({\it e.g.},
the $z$-axis), we project the spin four-vector in an invariant way
on the three directions in momentum space defined by the composite
four-vectors (\ref{TwoTwistorPRDef}) orthogonal to the four-vector
(\ref{TwoTwistorPDef}). We observe that using eq. (\ref{TijDef})
one can calculate the TPB of the $t^r$,
\begin{equation}\label{TAlgebra}
\{t^r,t^s\}=\epsilon^{rsu}t^u \quad .
\end{equation}
Thus, the  $t^r$ variables determine a $su(2)$ algebra of
invariant spin projections.

The composite four-momentum formulae (\ref{TwoTwistorPDef}) can be
supplemented
   with the  composite Lorentz
generators $M_{\alpha\beta},M_{\dalpha\dbeta}$ \bl
\begin{eqnarray}
M_{\alpha\beta}&=
&\pi_{(\alpha}\omega_{\beta)}+\eta_{(\alpha}\lambda_{\beta)}\quad,\label{MNoDotDef}\\
M_{\dalpha\dbeta}&=
&\opi_{(\dalpha}\oomega_{\dbeta)}+\oeta_{(\dalpha}\overline{\lambda}_{\dbeta)}\label{MDotDef}
\quad.
\end{eqnarray} \el
\noindent
As mentioned, the four-vector (\ref{RelationPLMomentum})
 can be identified with the composite Pauli-Luba\'{n}ski
$W_{\mu}$ $(W_\mu=\frac{1}{2}\epsilon_{\mu\nu\rho\tau}P^\nu
M^{\rho\tau})$ that defines the square of the relativistic spin
four-vector through $S^\mu S_\mu=S^2=\frac{W^2}{P^2}$ (so that
$W^2=-m^2\mathbf{s}^2 $). From (\ref{POrthNormBasis}) and
(\ref{RelationPLMomentum}) we obtain
\begin{equation}\label{S2}
S^2=-S^{(r)}S^{(r)}=-((t_1)^2+(t_2)^2+(t_3)^3) \quad .
\end{equation}
Therefore, the relativistic spin square is defined in any Lorentz
frame as the $su(2)$ Casimir $\mathbf{t}^2$,
 here given by the sum of the squares of the three
invariant spin projections.

In order to complete the description of spin in the two-twistor
space one can show that $(y^\adb= \mathrm{Im}\, z^\adb)$
\cite{Hu79,Tod77,Unknown-6}
\begin{equation}\label{yRelationWEP}
y^\adb=\frac{1}{P^2}\left(W^\adb-t^0 P^\adb\right)\quad,
\end{equation}
where
\begin{equation}\label{eDefTwist}
t^0=Z^A_{\phantom{A};i}\oZ_A^{\phantom{A};i}\quad .
\end{equation}
The fourth generator $t^0$ enlarges $su(2)$ to $u(2)$ due to the
vanishing of the TPB relation
\begin{equation}\label{PBRelT-e}
\{t^r,t^0\}=0\quad ;
\end{equation}
 $t^0$ can be associated with an internal $U(1)$
charge. Thus, we see from eq. (\ref{yRelationWEP}) that the
non-trivial Pauli-Luba\'{n}ski four-vector  and the internal
charge (\ref{eDefTwist}) generate the non-vanishing imaginary part
of the composite spacetime.

Following our earlier papers \cite{Bette04,Be05} we modify now the
standard Penrose formula (\ref{ComplexZFormula}) by introducing the
real composite spacetime coordinates $X_\mu$ firstly proposed by
Bette and Zakrzewski \cite{Be97},
\begin{equation}\label{XComDef}
X_\mu=x_\mu+\Delta x_\mu\quad ;
\end{equation}
they differ from $x_\mu$ by the shift $\Delta x_\mu$ (see eq.
(\ref{DeltaX}) below). In contrast with the non-commuting
variables $x^\mu$, eq. (\ref{XNonCommutative}), the variables
$X_\mu$ satisfy the conventional TPB as composites of
two-twistors, namely \bl
\begin{eqnarray}
\{X_\mu,X_\nu\}=\{P_\mu,P_\nu\}&=&0\label{XX-PPConvPB}\quad,\\
\{P_\mu,X_\nu\}&=&\eta_{\mu\nu}\label{XPConvPB}\quad.
\end{eqnarray}
\el \noindent
It follows from (\ref{XX-PPConvPB}) that after
quantizing the PB we will obtain a quantized theory with commuting
spacetime coordinates.

In the rest of the paper we proceed as follows. In Sec.
\ref{Dynamics} we discuss the explicit form of the modification
(\ref{XComDef}). Subsequently we define the two-twistor
counterpart of (\ref{OneTwistorRelPhaSpaPicture}) by introducing
the Liouville one-form on the 18-dimensional enlarged relativistic
phase space $\mathcal{M}^{18}$ \cite{Bette04}
\begin{equation}\label{M18}
\mathcal{M}^{18}=
(X_\mu,P_\mu,\eta_\alpha,\oeta_\dalpha,\sigma_\alpha,\osigma_\dalpha,e,\phi)\quad.
\end{equation}
In order to identify the manifold $\mathcal{M}^{18}$ with the
sixteen-dimensional two-twistor space (see (\ref{TwoTwistorDef}))
we have to introduce two constraints\footnote{In comparison with
\cite{Bette04} we employ here two equivalent, but different,
second class constraints.}. By expressing the Liouville one-form
(\ref{TwoTwistorLiouvilleInitial}) in terms of the variables of
$\mathcal{M}^{18}$ we find two second class constraints $R_2,R_3$
which encode the composite two-twistor structure of the variables
(\ref{M18}). These constraints are taken into consideration in the
quantization procedure by introducing suitable Dirac brackets.

In Sec. \ref{Sect.Quantization} we show  to what extent the Dirac
brackets in the space (\ref{M18}) can be correctly quantized {\it
i.e.}, without the violation of associativity. It turns out that
the quantized Dirac brackets can be only consistently  defined in
the subspace $\mathcal{\tilde{M}}^{14}=
(P_\mu,\eta_\alpha,\oeta_\dalpha,\sigma_\alpha,\osigma_\dalpha,e,\phi)$
of $\mathcal{M}^{18}$. We introduce further a differential
realization of the internal momenta
$(\sigma_\alpha,\osigma_\dalpha,e)$ in the generalized momentum
space $(P_\mu,\eta_\alpha,\oeta_\dalpha,\phi)$. In such a
generalized Schr\"{o}dinger representation the dynamics will be
described by four `physical' first class constraints determining
the mass, spin, invariant spin projection and electric charge.
They take the form of a mass shell condition and three
differential wave equations. We shall look for the explicit
solution of the wave equations describing spin in the form of a
power series
\begin{widetext}
\begin{equation}
\label{generalPsi}
 \Psi(P_\mu,\eta_\alpha,\oeta_\dalpha,\phi)=
\sum_{n,m=0}\sum_{\substack{(\alpha_1,\dots,\alpha_n)\\(\dbeta_1,\dots,\dbeta_m)}}
\eta_{\alpha_1}\cdots\eta_{\alpha_n}\oeta_{\dbeta_1}\cdots
\oeta_{\dbeta_m}\Psi^{(\alpha_1\dots\alpha_n)(\dbeta_1\dots\dbeta_m)}(P_\mu,\phi)\;,
\end{equation}
where the values of the four-momentum $P_\mu$ are restricted to
the mass-shell $P^2=m^2$ and the  dependence on the gauge variable
$\phi$ may be simply factorized out as a suitable $U(1)$ phase
factor. We shall express the higher spin fields in Minkowski space
by performing the standard Fourier transform
\begin{equation}\label{Fourier1}
{\Psi}^{(\alpha_1\cdots\alpha_n)(\dbeta_1\cdots\dbeta_m)}(\widetilde{x}_\mu)=
\frac{1}{(2\pi)^4}\int d^4P_\mu\ e^{iP_\mu
\widetilde{x}^\mu}\Psi^{(\alpha_1\cdots\alpha_n)(\dbeta_1\cdots\dbeta_m)}(P_\mu)
\quad .
\end{equation}
We stress here that the spacetime coordinates $\widetilde{x}_\mu$
appearing in the formula (\ref{Fourier1}) cannot be identified with
the commuting composite spacetime coordinates $X_\mu$ in
$\mathcal{M}^{18}$.

 We conclude in Sec. \ref{FinalRemarks} with some
remarks and open problems.
\end{widetext}

\section{From the two-twistor free model to a particle model
in enlarged spacetime} \label{Dynamics} \setcounter{equation}{0}

Let us modify the Penrose formula (\ref{ComplexZFormula}) in order
to introduce new complex composite spacetime coordinates $Z^\adb$
defined as follows
\begin{equation}\label{ZShift}
z^\adb\longrightarrow Z^\adb=z^\adb+\Delta z^\adb\quad,
\end{equation}
where
\begin{equation}\label{DeltaZ}
\Delta z^\adb=\Delta x^\adb+i\Delta y^\adb=
\frac{i\rho}{|f|^2}\pi^\alpha\oeta^\dbeta\quad.
\end{equation}
and $\rho=\frac{1}{2}(t_1-it_2)$. A calculation shows that \bl
\begin{eqnarray}
\Delta x_\adb &=
  &\frac{1}{2|f|^2}\left(t^1 P^2_{\adb}+t^2 P^1_\adb\right)\label{DeltaX}\quad,\\
\Delta y_\adb &=
  &\frac{1}{2|f|^2}\left(t^1P^1_\adb-t^2P^2_\adb\right)\label{DeltaY}\quad.
\end{eqnarray}
\el

We use now the non-linear change of twistor variables
$Z^A_{\phantom{A};i},\oZ_A^{\phantom{A};i}$ into the eighteen
coordinates of $\mathcal{M}^{18}$ given by the formula
(\ref{TwoTwistorPDef}) for $P_\mu$ and the following ones
\cite{Bette04,Be05} \bl
\begin{eqnarray}
X_\adb&=&\mathrm{Re}\,
Z_\adb=\frac{i}{2f}\left(\omega^\alpha\oeta^\dbeta-\lambda^\alpha\opi^\dbeta\right)
-\nonumber\\
&&-\frac{i}{2\of}\left(\eta^\alpha\oomega^\dbeta-\pi^\alpha\overline{\lambda}^\dbeta\right)+
\nonumber\\
&& +\frac{i}{2|f|^2}\rho\pi^\alpha\oeta^\dbeta-
  \frac{i}{2|f|^2}\orho\eta^\alpha\opi^\dbeta\quad,\label{XDef}\\
\sigma^\alpha&=&-\frac{1}{\of}\left(\orho\eta^\alpha+t_3\pi^\alpha\right)\quad,\label{SigmaDef}\\
\osigma^\dalpha&=&-\frac{1}{f}\left(\rho\oeta^\dalpha+t_3\opi^\dalpha\right)\quad,\label{oSigmaDef}\\
e&=&t_0+t_3\label{RelE-t3}\quad,\\
\phi&=&\frac{i}{2}\ln \frac{\of}{f}\quad.\label{phiDef}
\end{eqnarray}
\el

Using (\ref{TwoTwistorPDef})-(\ref{P2}) and (\ref{phiDef}) we
obtain\bl
\begin{eqnarray}
f&=&\sqrt{\frac{P^2}{2}}e^{i\phi}\quad,\label{fDefWithPhi}\\
\pi_\alpha&=&-\frac{1}{f}P_{\adb}\oeta^\dbeta\quad,\label{piExplicit}\\
\opi_\dalpha&=&-\frac{1}{\of}\eta^\beta P_{\beta\dalpha}\quad.\label{opiExplicit}
\end{eqnarray}
\el \noindent It follows from eqs.
(\ref{SigmaDef})-(\ref{oSigmaDef}) that the invariant spin
components $t_1, t_2$ and $t_3$ (eqs. (\ref{TijDef}), (\ref{S-R}))
are given in terms of the variables $t_3, \rho, \orho$ by the
formulae \bl
\begin{eqnarray}
t^3&=&\eta^\alpha\sigma_\alpha=
\oeta^\dalpha\osigma_\dalpha\quad,\label{t3Ireal}\\
\rho=t_1-it_2&=&\opi_\dalpha\osigma^\dalpha=-\frac{1}{\of}\eta_\alpha
P^{\adb}\osigma_\dbeta\label{rhoExplicit}\quad,\\
\orho=t_1+it_2&=&\pi_\alpha\sigma^\alpha=-\frac{1}{f}\sigma_\alpha
P^\adb \oeta_\dbeta\quad.\label{orhoExplicit}
\end{eqnarray}
\el
  Further, the linear combination (\ref{RelE-t3}) of the
internal scalar charge $t^0$ and $t^3$ will be called electric
charge. Indeed, if we identify $t^0$ with $\frac{Y}{2}$
(`hypercharge') and $t^3$ with the third isospin component, eq.
(\ref{RelE-t3}) takes  the form of the Gell-Mann-Nishijima formula
for the electric charge\footnote{The identification of $e$ as the
electric charge can be ultimately justified by coupling our model
to the electromagnetic field, which is outside of the scope of
this paper.}.

Using the formulae (\ref{TwoTwistorPDef}),
(\ref{XDef})-(\ref{phiDef}), we can now re-express the free
two-twistor Liouville one-form (\ref{TwoTwistorLiouvilleInitial})
as
\begin{equation}\label{TwoTwistorLiouvilleM18}
\Theta^{(2)}=P_\mu dX^\mu-i\left(\osigma^\dalpha d\oeta_\dalpha-\sigma^\alpha
d\eta_\alpha\right)+ed\phi\quad,
\end{equation}
where the 18 coordinates of $\mathcal{M}^{18}$ are restricted by
two constraints. From  definitions (\ref{TwoTwistorPDef}) and
(\ref{XDef})-(\ref{phiDef}) we can obtain the following
constraints:

\bl
\begin{eqnarray}
R_1&=&\sigma_\alpha P^{\adb}\osigma_\dbeta-\mathbf{t}^2=0\quad
(\mathbf{t}^2=\rho\orho +
(t^3)^2)\ ,\quad\label{Constraint-sPos-t2}\\
R_2&=&\eta_\alpha P^\adb \oeta_\dbeta-\frac{1}{2}P^2=0\quad.
\label{Constraint-ePoe-1/2P2}
\end{eqnarray}
\el

\noindent The relation (\ref{t3Ireal}) implies additionally
the reality constraint
\begin{equation}\label{Constraint-es-oeos}
R_3=\eta^\alpha\sigma_\alpha-\oeta^\dalpha\osigma_\dalpha=0\quad.
\end{equation}

It can be shown that only two constraints out of the three $R_1,
R_2, R_3$ are independent. Indeed, if we introduce
\begin{equation}
\widetilde{R}_1\equiv R_1+\frac{2\sigma^\alpha P_\ada
\osigma^\dalpha}{P^2}R_2=\frac{2\sigma^\alpha
P_\ada\osigma^\dalpha \eta_\beta
P^{\beta\dbeta}\oeta_\dbeta}{P^2}-\mathbf{t}^2
\end{equation}
by using the definition of $\mathbf{t}^2$ (see
(\ref{t3Ireal})-(\ref{orhoExplicit})) and then use the following
Fierz identity
\begin{eqnarray}
 (\sigma^\alpha \sigma^{\mu}_\ada \osigma^\dalpha)
(\eta^\beta\sigma^{\nu}_{\beta\dbeta}\oeta^\dbeta)&=&
-(\sigma^\alpha\eta_\alpha)(\oeta^\dbeta(\tilde{\sigma}^\nu\sigma^\mu)_{\dbeta\dalpha}\osigma^\dalpha)
+\nonumber\\
&&+(\sigma^\alpha\sigma^\nu_{\alpha\dbeta}\oeta^\dbeta)(\eta^\beta
\sigma^\mu_{\beta\dalpha}\osigma^\dbeta)\label{Fierz}
\end{eqnarray}
one can show that $\widetilde{R}_1\equiv 0$, {\it i.e.}
$R_1=0$ follows from $R_2=0$.

We stress that the two algebraic constraints $R_2=R_3=0$ encode the
composite structure of the variables (\ref{M18}) as expressed in
terms of primary twistor variables (\ref{TwoTwistorDef}).

\begin{widetext}
The Liouville one-form (\ref{TwoTwistorLiouvilleM18}) and the
constraints $R_2=R_3=0$ provide the following action describing a
relativistic particle model $(\dot{a}\equiv\frac{da}{d\tau})$
\begin{equation}\label{Action}
S=\int d\tau L(\tau)=\int d\tau \left[P_\mu\dot{X}^\mu -
i\left(\osigma^\dalpha\dot{\oeta}_\dalpha-\sigma^\alpha\dot{\eta}_\alpha\right)+e\dot{\phi}+\lambda_2
R_2+\lambda_3 R_3\right]\;,
\end{equation}
This action, supplemented only with the algebraic constraints on
$\mathcal{M}^{18}$, can be quantized canonically.  The action leads
to the canonical PB (CPB) (\ref{XX-PPConvPB})-(\ref{XPConvPB}) in
the spacetime sector of $\mathcal{M}^{18}$ plus the following
(non-vanishing) remaining ones

\bl
\begin{eqnarray}
\{\eta_\alpha,\sigma^\beta\}_C&=&i\delta_\alpha^\beta\quad,\label{CPBetasigma}\\
\{\oeta_\dalpha,\osigma^\dbeta\}_C&=&-i\delta_\dalpha^\dbeta\quad,\label{CPBoetaosigma}\\
\{e,\phi\}_C&=&1\quad,\label{CPBephi}
\end{eqnarray}
\el \noindent where $\{\ ,\ \}_C$ means canonical PB.

Before considering the dynamical first class constraints that
 fix the values of the mass, spin, spin projection and
electric charge, we have to introduce Dirac brackets consistent
with the algebraic second class constraints $R_2,R_3$. Since
\begin{equation}
\label{CB-R2-R3}
 \{R_2,R_3\}_C=-2i\eta_\alpha P^\ada \oeta_\dalpha=-iP^2 \quad,
\end{equation}
the Dirac bracket of two dynamical quantities $A,B$ takes the form
\begin{equation}\label{Dirac-Bracket}
\{A,B\}_D=\{A,B\}_C-\left[\{A,R_2\}_C\frac{1}{iP^2}\{R_3,B\}_C-
           \{A,R_3\}_C\frac{1}{iP^2}\{R_2,B\}_C\right] \; .
\end{equation}
\end{widetext}
Thus, within the generalized phase space $\mathcal{M}^{18}$, the
only coordinates $Y$ that possess non-vanishing Poisson brackets
with the constraints $R_2$ and $R_3$ are
\begin{eqnarray}
\{Y,R_2\}_C&\neq& 0 \qquad \text{if}\quad Y=X_\ada,\sigma_\alpha,\osigma_\dalpha\\
\{Y',R_3\}_C&\neq& 0 \qquad \text{if}\quad
Y'=\eta_\alpha,\oeta_\dalpha,\sigma_\alpha,\osigma_\dalpha \quad .
\end{eqnarray}
Hence, according to (\ref{Dirac-Bracket}), only the Dirac
$\{Y,Y'\}_D$ brackets will be different from the canonical
$\{Y,Y'\}_C$ ones. This leads  to two sets of non-zero
Dirac brackets, one involving the spacetime variables $X_\ada$,
\bl
\begin{eqnarray}
\{X_{\ada},\eta_\gamma\}_D&=
  &-\frac{1}{P^2}(P_\ada-\eta_\alpha\oeta_\dalpha)\eta_\gamma\quad,\label{DB-X-eta}\\
\{X_{\ada},\oeta_\dgamma\}_D&=
  &-\frac{1}{P^2}(P_\ada-\eta_\alpha\oeta_\dalpha)\oeta_\dgamma\quad,\\
\{X_{\ada},\sigma^\gamma\}_D&=
  &\frac{1}{P^2}(P_\ada-\eta_\alpha\oeta_\dalpha)\sigma^\gamma\quad,\\
\{X_{\ada},\osigma^\dgamma\}_D&=
  &\frac{1}{P^2}(P_\ada-\eta_\alpha\oeta_\dalpha)\osigma^\dgamma\quad,\label{DB-X-osigma}
\end{eqnarray}
and another  involving only those of the spin sector,
\begin{eqnarray}
\{\eta_\alpha,\sigma^\beta\}_D&=
  &i\delta_\alpha^\beta-\frac{i}{P^2}\eta_\alpha P^{\beta\dgamma}\oeta_\dgamma\quad,\label{DB-eta-sigma}\\
\{\eta_\alpha,\osigma^\dbeta\}_D&=
  &\frac{i}{P^2}\eta_\alpha\eta_\gamma P^{\gamma\dbeta}\quad,\\
\{\sigma^\alpha,\sigma^\beta\}_D&=
  &\frac{i}{P^2}\sigma^{[\alpha}P^{\beta]\dgamma}\oeta_\dgamma\quad,\\
\{\sigma^\alpha,\osigma^\dbeta\}_D&=
  &-\frac{i}{P^2}\left(\eta_\gamma P^{\gamma\dbeta}\sigma^\alpha+
  P^{\alpha\dgamma}\oeta_\dgamma\osigma^\dalpha\right)\label{DB-sigma-osigma}\quad,\\
\{\oeta_\dalpha,\sigma^\beta\}_D&=
  &-\frac{i}{P^2}P^{\beta\dgamma}\oeta_\dgamma\oeta_\dalpha\quad,\\
  \{\oeta_\dalpha,\osigma^\dbeta\}_D&=&-i\delta^\dbeta_\dalpha+
  \frac{i}{P^2}\oeta_\dalpha\eta_\gamma P^{\gamma\dbeta}\quad,\\
\{\osigma^\dalpha,\osigma^\dbeta\}_D&=
  &-\frac{i}{P^2}\eta_\gamma
P^{\gamma[\dbeta}\osigma^{\dalpha]}\quad \, .\label{DB-osigma-osigma}
\end{eqnarray}\el
All other  Dirac brackets in $\mathcal{M}^{18}$ coincide with the
canonical ones.

If we now replace the relations
(\ref{DB-X-eta})-(\ref{DB-osigma-osigma}) by their quantum
analogues,
\begin{equation}
\label{From-DB-to-Quantum}
  a\rightarrow \widehat{a}\quad,\qquad\{a,b\}_D\rightarrow \frac{1}{i}\left[\widehat a,\widehat
  b\right] \quad ,
\end{equation}
 and provide a rule for the ordering of variables
associated with the non-commuting operators, one can show that only
the seven relations (\ref{DB-eta-sigma})-(\ref{DB-osigma-osigma})
above produce a consistent ({\it i.e.}, satisfying the Jacobi
identity) set of commutators \bl
\begin{eqnarray}
\phantom{}[\weta_\alpha,\wsigma^\beta]&=&-\delta_\alpha^\beta+
  \frac{1}{P^2}\weta_\alpha \wP^{\beta\dgamma}\woeta_\dgamma\quad,\label{QB-eta-sigma}\\
\phantom{}[\weta_\alpha,\wosigma^\dbeta]&=
  &-\frac{1}{P^2}\weta_\alpha\weta_\gamma \wP^{\gamma\dbeta}\quad,\\
\phantom{}[\wsigma^\alpha,\wsigma^\beta]&=
  &-\frac{1}{P^2}\wsigma^{[\alpha}\wP^{\beta]\dgamma}\woeta_\dgamma\quad,\\
\phantom{}[\wsigma^\alpha,\wosigma^\dbeta]&=
  &\frac{1}{P^2}\left(\weta_\gamma \wP^{\gamma\dbeta}\wsigma^\alpha+
  \wP^{\alpha\dgamma}\woeta_\dgamma\wosigma^\dalpha\right)\quad,\label{QB-sigma-osigma}\\
\phantom{}[\woeta_\dalpha,\wsigma^\beta]&=
  &\frac{1}{P^2}\wP^{\beta\dgamma}\woeta_\dgamma\woeta_\dalpha\quad,\\
\phantom{}[\woeta_\dalpha,\wosigma^\dbeta]&=
  &\delta^\dbeta_\dalpha-\frac{1}{P^2}\woeta_\dalpha\weta_\gamma \wP^{\gamma\dbeta}\quad,\\
\phantom{}[\wosigma^\dalpha,\osigma^\dbeta]&=
  &\frac{1}{P^2}\weta_\gamma
\wP^{\gamma[\dbeta}\wosigma^{\dalpha]}\quad.\label{QB-osigma-osgima}
\end{eqnarray}\el
In particular, we note that the quantization
(\ref{QB-sigma-osigma}) of the Dirac bracket
(\ref{DB-sigma-osigma}) requires the specific order of the
operators indicated by the {\it r.h.s.} of formula
(\ref{DB-sigma-osigma}). Unfortunately, for any choice of
ordering, the quantization of the Dirac brackets
(\ref{DB-X-eta})-(\ref{DB-X-osigma}) involving the spacetime
coordinate $X_\mu$ leads to a non-associative algebra\footnote{The
difficulties with the fulfillment of the Jacobi identity (JI) by
the quantized Dirac brackets are well-known (see {\it e.g.}
\cite{Cuxx}).}.

The algebra (\ref{QB-eta-sigma})-(\ref{QB-osigma-osgima}) can be
realized in terms of differential  operators on the following
8-dimensional enlarged momentum space
\begin{equation}\label{P8}
\mathcal{P}^8=\mathcal{P}_k=(P_\mu,\eta_\alpha,\oeta_\dalpha,\phi;R_2=0)
\quad .
\end{equation}
One can show that the operators
$\wsigma^\alpha,\wosigma^\dalpha,\widehat{e}$ have the following
generalized Schr\"{o}dinger realization\footnote{Notice that we do not need to
check algebraically all the JI's for
(\ref{QB-eta-sigma})-(\ref{QB-osigma-osgima}); it is sufficient to
show that in differential realization ($\weta_\alpha=\eta_\alpha,\
\woeta_\dalpha=\oeta_\dalpha,\ \widehat{P}^{\ada}=P^\ada$ and
 formulae
(\ref{Diff-Reali-sigma})-(\ref{Diff-Reali-e})) the Dirac bracket algebra
(\ref{QB-eta-sigma})-(\ref{QB-osigma-osgima}) is satisfied.}

\begin{eqnarray}
\widehat{\sigma}^\alpha&=&\frac{\partial}{\partial
\eta_\alpha}-P^{\alpha\dbeta}\oeta_\dbeta\frac{1}{P^2}\left(\eta_\gamma\frac{\partial}{\partial
\eta_\gamma}+\oeta_\dgamma\frac{\partial}{\partial
\oeta_\dgamma}\right)\quad,\quad\label{Diff-Reali-sigma}\\
\widehat{\osigma}^\dalpha&=&-\frac{\partial}{\partial
\oeta_\dalpha}+\eta_\beta
P^{\beta\dalpha}\frac{1}{P^2}\left(\eta_\gamma\frac{\partial}{\partial
\eta_\gamma}+\oeta_\dgamma\frac{\partial}{\partial
\oeta_\dgamma}\right)\ ,\quad\label{Diff-Reali-osigma}\\
\widehat{e}&=&\frac{1}{i}\frac{\partial}{\partial \phi} \quad .\quad
\label{Diff-Reali-e}
\end{eqnarray}
These  operators satisfy
(\ref{QB-eta-sigma})-(\ref{QB-osigma-osgima}), to which we may add
$[\widehat{\phi},\widehat{e}]=i$. It can be checked that in the
differential realization
(\ref{Diff-Reali-sigma})-(\ref{Diff-Reali-e}) the constraint
$R_3=0$ is identically satisfied.

In order to determine the relativistic particle states with
 definite mass $m$, spin $s$, invariant spin projection
  $s_3$ and electric charge $e_0$, we supplement the action
(\ref{Action}) with the following physical constraints: \bl
\begin{eqnarray}
R_4&=&\mathbf{t}^2-s(s+1)=0\quad,\label{Constraint-t2-s(s+1)}\\
R_5&=&t_3-s_3=0\quad,\label{Constraint-t3-s3}\\
R_6&=&P^2-m^2=0\quad,\label{Constraint-P2-m2}\\
R_7&=&e-e_0=0\label{Constraint-e-e0} \quad .
\end{eqnarray}
\el
It can be shown that the CPB of the constraints
$R_2,\dots,R_7$ provide the following  canonical Poisson brackets
algebra besides (\ref{CB-R2-R3}): \bl
\begin{eqnarray}
\{R_2,R_4\}_C&=&-i\eta_\alpha P^\ada\oeta_\dalpha R_3\label{ConstraintPB24}\\
\{R_2,R_k\}_C&=&0\quad\qquad k=5,6,7\label{ConstraintPB2k}\\
\{R_n,R_m\}_C&=&0\quad\qquad n,m=3,\dots,7\label{ConstraintPBmn}
\end{eqnarray}
\el The set of relations
(\ref{ConstraintPB24})-(\ref{ConstraintPBmn}) shows that the
constraints $R_4,R_5,R_6,R_7$ are first class.

\section{First-quantized theory and wave equations}
\label{Sect.Quantization} \setcounter{equation}{0} The quantum form
of the first class constraints
(\ref{Constraint-t2-s(s+1)})-(\ref{Constraint-e-e0}) provides the
following four wave equations for the function
$\Psi(\mathcal{P}_k)=\Psi(P_\mu,\eta_\alpha,\oeta_\dalpha,\phi)$ (we
set $\hbar=c=1$)

\bl
\begin{eqnarray}
R_4=0:&&
\left[\widehat{t_3}^2+\widehat{\rho\orho}-s(s+1)\right]\Psi(\mathcal{P}_k)=0\quad,
 \label{DiffReal-R4}\\
R_5=0:&&\left[\widehat{t_3}-s_3\right]\Psi(\mathcal{P}_k)=0\quad,\label{DiffReal-R5}\\
R_6=0:&& \left[P^2-m^2\right]\Psi(\mathcal{P}_k)=0\quad,\label{DiffReal-R6}\\
R_7=0:&& \left[\frac{\partial}{\partial
\phi}+ie_0\right]\Psi(\mathcal{P}_k)=0\quad,\label{DiffReal-R7}
\end{eqnarray}
\el where \bl
\begin{eqnarray}
\widehat{t_3}&=&\frac{1}{2}\left(\oeta_\dalpha\frac{\partial}{\partial
\oeta_\dalpha}-\eta_\alpha\frac{\partial}{\partial
\eta_\alpha}\right)\quad,\label{Operator-k}\\
\widehat{\rho}&=&\sqrt{\frac{2}{P^2}}e^{i\phi}\eta^\alpha P_\ada
\frac{\partial}{\partial
\oeta_\dalpha}\quad,\label{Operator-rho}\\
\widehat{\orho}&=&-\sqrt{\frac{2}{P^2}}e^{-i\phi}\oeta^\dalpha
P_\ada \frac{\partial}{\partial
\eta_\alpha}\quad,\label{Operator-orho}\\
\widehat{\rho\orho}&=&\frac{1}{2}\left(\widehat{\rho}\widehat{\orho}+
  \widehat{\orho}\widehat{\rho}\right) \quad ,
\end{eqnarray}
\el\noindent
 and the generators $\widehat{t_3}$,
$\;\widehat{\rho}=\widehat{t}_1-i\widehat{t}_2$ and
$\widehat{\orho}=\widehat{t}_1+i\widehat{t}_2$ satisfy the
commutation relations of the $su(2)$ algebra,
\begin{equation}
[\wrho,\worho]=-2t_3\quad .
\end{equation}

A general solution of eq. (\ref{DiffReal-R4})-(\ref{DiffReal-R7})
 can be written in the form described by the formula (1.30),
 where the bracketed indices are symmetric. The $\phi$-dependence
is determined by eq. (\ref{DiffReal-R7}) which gives
\begin{equation}\label{Sol.e_0}
\psi^{(\alpha_1\cdots\alpha_n)(\dbeta_1\cdots\dbeta_m)}(P_\mu,\phi)=
 e^{-ie_0\phi}\psi^{(\alpha_1\cdots\alpha_n)(\dbeta_1\cdots\dbeta_m)}(P_\mu)\quad,
\end{equation}
where $e_0$ is the (electric) charge associated to the $U(1)$
gauge transformation generated by the operator $\widehat{e}$ of
eq. (\ref{Diff-Reali-e}).  The standard complex conjugation
properties for $\Psi(\mathcal{P}_k)$ lead us to assume further the
following condition
\begin{equation}\label{Psi-real}
\psi^{(\alpha_1\cdots\alpha_n)(\dbeta_1\cdots\dbeta_m)}(P_\mu)=
\left(\psi^{(\beta_1\cdots\beta_m)(\dalpha_1\cdots\dalpha_n)}(P_\mu)\right)^*\quad.
\end{equation}

The action of the operators (\ref{Operator-k}),
(\ref{Operator-rho}) and (\ref{Operator-orho}) on the variables
$\eta_\alpha$ and $\oeta_\dalpha$ is given by \bl
\begin{eqnarray}
\label{spin3}
 \widehat{t_3}\, \eta_\alpha=
 -\frac{1}{2}\eta_\alpha&\qquad& \widehat{t_3}\, \oeta_\dalpha=\frac{1}{2}\oeta_\dalpha\\
\wrho\, \eta_\alpha=0 &\quad&\wrho\,\oeta_\dalpha
 =\sqrt{\frac{2}{P^2}}e^{i\phi}\eta^\alpha
P_\ada\qquad\quad \\
\worho\, \eta_\alpha=
 -\sqrt{\frac{2}{P^2}}e^{-i\phi}\oeta^\dalpha
P_\ada&\quad&\worho\, \oeta_\dalpha =0
\end{eqnarray}
\el

We now show that suitable polynomials in $\eta_\alpha$ and
$\oeta_\dalpha$ in eq. (\ref{generalPsi}), together with the
proper $P_\mu$ dependence of $\Psi$, describe states with definite
values of spin and spin projection, {\it i.e.} $|s,s_3>$,
satisfying (\ref{DiffReal-R4})-(\ref{DiffReal-R5}).  First, we see
that the $\eta_\alpha,\ \oeta_\dalpha$ variables themselves
correspond to $|1/2,-1/2>$, $|1/2,1/2>$, respectively, due to \bl
\begin{eqnarray}
\wtt \, \eta_\alpha&=&\frac{3}{4}\eta_\alpha\quad,\\
\wtt \, \oeta_\dalpha&=&\frac{3}{4}\oeta_\dalpha\quad.
\end{eqnarray}
\el
 and (\ref{spin3}). In general,
 \bl
\begin{eqnarray}
\wt3\,
\eta_{\alpha_1}\cdots\eta_{\alpha_n}\oeta_{\dalpha_1}\cdots\oeta_{\dalpha_m}
&=&
\frac{m-n}{2}\eta_{\alpha_1}\cdots\eta_{\alpha_n}\oeta_{\dalpha_1}
   \cdots\oeta_{\dalpha_m},\qquad\ \label{wt3-neta,moeta}\\
\wtt\,\eta_{\alpha_1}\cdots\eta_{\alpha_n}&=
   &\frac{n(n+2)}{4}\eta_{\alpha_1}\cdots\eta_{\alpha_n}\quad,\label{wtt-neta}\\
\wtt\,\oeta_{\dalpha_1}\cdots\oeta_{\dalpha_m}&=
  &\frac{m(m+2)}{4}\oeta_{\dalpha_1}\cdots\oeta_{\dalpha_m}\label{wtt-noeta}
  \quad ,
\end{eqnarray}\el
\noindent
where we recall that the Lorentz scalar $\wt3$ describes
the projection of the relativistic spin four-vector $W_\mu$ on the
direction $P^{(3)}_\mu$, orthogonal to
 $P_\mu$ (see (\ref{TwoTwistorPRDef}) and (\ref{S-R})). In order to
 identify the relations
(\ref{wt3-neta,moeta})-(\ref{wtt-noeta}) with
(\ref{DiffReal-R4})-(\ref{DiffReal-R5}) we set
\begin{equation}
s=\frac{n}{2}\quad \text{or} \quad s=\frac{m}{2}\quad,
\end{equation} and
\begin{equation}
s_3=\frac{m-n}{2}\quad.
\end{equation}
{}From eqs. (\ref{wt3-neta,moeta})-(\ref{wtt-noeta}) we conclude
 that
$\eta_{\alpha_1}\cdots\eta_{\alpha_n}\sim|\frac{n}{2},-\frac{n}{2}>$
 and $\oeta_{\dalpha_1}\cdots\oeta_{\dalpha_n}$ $\sim|\frac{n}{2},\frac{n}{2}>$.

The general theory of $su(2)$ representations tells us that one can
get all the states starting either from the lowest one, $|s,-s>$,
and acting with the raising operator $\worho$ or from the highest
one, $|s,s>$, and acting with the lowering operator $\wrho$. The
action of the ladder operators on a product of $\eta_\alpha$'s or
$\oeta_\dalpha$'s can be summarized as follows,
\begin{widetext}\bl
\begin{eqnarray}
\left|\frac{n}{2},\frac{m-n}{2}\right>_{{\alpha_1}\cdots{\alpha_n}}&\sim
  &(\,\worho\,)^m\eta_{\alpha_1}\cdots\eta_{\alpha_n}\nonumber\\
&=&m!\left(\sqrt{\frac{2}{P^2}}\right)^m
e^{-im\phi}\cdot\nonumber\\
&&\cdot\sum_{c\in\mathcal{C}^m_n}
\sum_{\dbeta_1\cdots\dbeta_m}P_{\alpha_1}^{\phantom{\alpha_1}\dbeta_1}\cdots
P_{\alpha_m}^{\phantom{\alpha_m}\dbeta_m}\oeta_{\dbeta_1}
  \cdots\oeta_{\dbeta_m}\eta_{\alpha_{m+1}}\cdots\eta_{\alpha_n}\nonumber\quad,\\\label{m-worho}
\end{eqnarray}
\begin{eqnarray}
  \left|\frac{n}{2},\frac{n-m}{2}\right>_{{\dbeta_1}
  \cdots{\dbeta_n}}&\sim&(\,\wrho\,)^m\oeta_{\dbeta_1}\cdots\oeta_{\dbeta_n}\nonumber\\
&=&(-1)^{m}m!\left(\sqrt{\frac{2}{P^2}}\right)^m
e^{im\phi}\cdot\nonumber\\
&&\cdot\sum_{c\in\mathcal{C}^m_n}
\sum_{\alpha_1\cdots\alpha_m}P^{\alpha_1}_{\phantom{\alpha_1}\dbeta_1}\cdots
P^{\alpha_m}_{\phantom{\alpha_m}\dbeta_m}\eta_{\alpha_1}
  \cdots\eta_{\alpha_m}\oeta_{\dbeta_{m+1}}\cdots\oeta_{\dbeta_n}\nonumber\quad.\\
  \label{m-wrho}
\end{eqnarray}\el
In particular if $m>n$ both expressions are zero, {\it i.e.} we
generate by the method described by formulae
(\ref{m-worho})-(\ref{m-wrho}) the $(2s+1)$-dimensional basis (we
recall that $s=\frac{n}{2}$) of the irreducible representation with
spin $s$.  We may thus
describe a wavefunction
 with definite spin $s$ and spin projection $s_3$ $(-s\le s_3\le s)$
 by starting from the lowest value of $s_3$ (see (\ref{m-worho})) or
 from the highest one (see (\ref{m-wrho})).
 We have the following two sequences generating the
  representation space basis
 \bl
\begin{eqnarray}
\wrho\  |s,s> \rightarrow |s,s-1>  &\dots & \wrho\ |s,-s+1>
\rightarrow
|s,-s>\label{wrho ladder}\\
\worho\  |s,-s> \rightarrow |s,-s+1>  &\dots & \worho\ |s,s-1>
\rightarrow |s,s>\label{worho ladder}
\end{eqnarray}
\el\noindent
 For example, for spin $s=1$ we have the following
  three-dimensional
basis, generated by the sequence (\ref{m-worho}) \bl
\begin{eqnarray}
\left| 1,-1\right>_{\alpha\beta} & =
&\eta_\alpha\eta_\beta\label{Spin-1,-1}\\
\left|1,0\right>_{\alpha\beta} &= & \sqrt{\frac{2}{P^2}}e^{-i\phi}
\left(P_\alpha^{\phantom{\alpha}\dalpha}\oeta_\dalpha\eta_\beta+
  P_\beta^{\phantom{\alpha}\dalpha}\oeta_\dalpha\eta_\alpha\right)
  \label{Spin-1,0}\\
\left| 1,1\right>_{\alpha\beta} & =
&\frac{4}{P^2}e^{-2i\phi}P_\alpha^{\phantom{\alpha}\dalpha}P_\beta^{\phantom{\alpha}\beta}
 \oeta_\dalpha\oeta_\dbeta\label{Spin-1,1}
\end{eqnarray} \el
or, alternatively, by the sequence (\ref{m-wrho}),

\bl
\begin{eqnarray}\label{11}
\left| 1,1\right>_{\dalpha\dbeta} & = &\oeta_\dalpha\oeta_\dbeta\label{Spin-1,1:2}\\
\left|1,0\right>_{\dalpha\dbeta} &= &
-\sqrt{\frac{2}{P^2}}e^{i\phi}\left(P^\alpha_{\phantom{\alpha}\dalpha}\eta_\alpha\oeta_\dbeta+
P^\alpha_{\phantom{\alpha}\dbeta}\eta_\alpha\oeta_\dalpha\right)\label{Spin-1,0:2}\\
\left| 1,-1\right>_{\dalpha\dbeta} & =
  & \frac{4}{P^2}e^{2i\phi} P^\alpha_{\phantom{\alpha}\dalpha}
  P^\beta_{\phantom{\alpha}\beta}\eta_\alpha\eta_\beta\label{Spin-1,-1:2} \quad .
\end{eqnarray} \el
\end{widetext}
We add here that the particular momentum dependent coefficients
(see  (\ref{Spin-1,0})--(\ref{Spin-1,1})
 and (\ref{Spin-1,0:2})--(\ref{Spin-1,-1:2}))
 are due to our definitions of the raising and lowering
 operators $\wrho$, $\worho$ which shift by one the
 invariant projections $s_3$ given by $\wt3$.

We now derive  in our framework the linear field equations for any
spin (Dirac for $s=\frac{1}{2}$ and Bargmann-Wigner for arbitrary
spin). Further, we consistently assume that the relations
(\ref{Psi-real}) are fulfilled. For $s=\frac{1}{2}$, {\it i.e.}
$n+m=1$ in (\ref{generalPsi}), the wavefunction is described by
one complex Weyl spinor
$\psi^\alpha(P_\mu)=(\psi^\dalpha(P_\mu))^*$ or equivalently by a
real four component Majorana spinor. The Weyl spinor
$\psi^\alpha(P_\mu)$ satisfies the Klein-Gordon equation (see
(\ref{DiffReal-R6}))
\begin{equation}\label{Klein-Gordon}
(P^2-m^2)\psi^\alpha(P_\mu)=0
\end{equation}
In order to introduce the Dirac equation we linearize the equation
(\ref{Klein-Gordon}) by introducing a new Weyl spinor
$\chi_\dbeta(P_\mu)$ by means of the defining equation (we assume
$m\neq 0$)
\begin{equation}\label{Def-chi}
\psi^\alpha(P_\mu)P_{\adb}=m\chi_\dbeta(P_\mu)
\end{equation}
{}From (\ref{Klein-Gordon}) and (\ref{Def-chi}) follows also that
\begin{equation}\label{Def-chi inv}
P^{\ada}\chi_\dalpha(P_\mu)=m\psi^\alpha(P_\mu)
\end{equation}
 The set of equations (\ref{Def-chi}), (\ref{Def-chi
inv}) provide the standard  momentum space Dirac equation in terms
of two Weyl equations coupled through the particle's mass.

In order to discuss the arbitrary spin case we introduce
 besides $\Psi(\mathcal{P}_k)$ a second function
$\Upsilon(\mathcal{P}_k)$ satisfying the relation
\begin{equation}\label{new6}
P_\ada\frac{\partial}{\partial\oeta_\dalpha}\Psi(\mathcal{P}_k)=
  m\frac{\partial}{\partial\eta^\alpha}\Upsilon(\mathcal{P}_k) \quad .
\end{equation}
Using (\ref{DiffReal-R4}) one gets from (\ref{new6})
\begin{equation}\label{new7}
P^\ada\frac{\partial}{\partial\eta^\alpha}\Upsilon(\mathcal{P}_k)=
  m\frac{\partial}{\partial\oeta_\dalpha}\Psi(\mathcal{P}_k)\quad .
\end{equation}
Equations (\ref{new6}) and (\ref{new7}) describe the generalized
Dirac equations in the enlarged momentum space (\ref{P8}).

Expressing (\ref{new6}), (\ref{new7}) in powers of $\eta_\alpha$
and $\oeta_\dalpha$ we obtain  from the linear terms the Dirac
equation (see eqs. (\ref{Def-chi}), (\ref{Def-chi inv})) and, from
the higher order terms, the Fierz-Pauli/Bargmann-Wigner equations
for arbitrary spin \cite{Unknown-17}\bl
\begin{eqnarray}
\psi^{(\alpha_1\cdots\alpha_k)(\dbeta_2\cdots\dbeta_l)}(P_\mu)P_{\alpha_1}^{\phantom{\alpha_1}\dbeta_1}&=
  &m\chi^{(\alpha_2\cdots\alpha_k)(\dbeta_1\cdots\dbeta_l)}(P_\mu)\quad,\nonumber\\&&\\
P^{\alpha_1}_{\phantom{\alpha_1}\dbeta_1}\chi^{(\alpha_2\cdots\alpha_k)(\dbeta_1\cdots\dbeta_l)}(P_\mu)&=
  &m\psi^{(\alpha_1\cdots\alpha_k)(\dbeta_2\cdots\dbeta_l)}(P_\mu)\quad.\nonumber\\&&
\end{eqnarray}\el

Finally, let us  consider the question of the spacetime picture in
our formalism. The equations
(\ref{DiffReal-R4})-(\ref{DiffReal-R7}) as well as (\ref{new6}),
(\ref{new7}) have been derived in the enlarged momentum space
(\ref{P8}). As we have shown in Sec. \ref{Dynamics}, the composite
real spacetime coordinates (\ref{XDef}) do commute with themselves
(see (\ref{XX-PPConvPB})), but they do not commute with the spin
sector variables (see (\ref{DB-X-eta})-(\ref{DB-X-osigma})). This
non-commutativity as derived from our model (\ref{Action}) is
reflected at the classical level
 by the non-vanishing  Dirac brackets. Unfortunately, we have not been able
 to quantize the  set of Dirac brackets which contain the
 composite space-time coordinates
 in a way that leads to an associative algebra {\it i.e.},
 satisfying Jacobi identities. At this stage of the development
 of our framework we have to abandon the idea that
 the commuting composite spacetime
coordinates of eq. (\ref{XDef})  describe `physical' spacetime.
Instead, we may introduce another set of commuting spacetime
coordinates by means of the standard Fourier transform in the
physical four-momentum sector of the manifold $\mathcal{P}_k$,
\begin{equation}\label{Fourier2}
{\Psi}(\widetilde{x}_\mu,\eta_\alpha,\oeta_\dalpha,\phi)=
\frac{1}{(2\pi)^4}\int d^4P_\mu\ e^{iP_\mu
\widetilde{x}^\mu}\Psi(P_\mu,\eta_\alpha,\oeta_\dalpha,\phi)\quad,
\end{equation}
which leads to the formula (1.31).
 Due to the constraint (\ref{DiffReal-R6}) the mass shell
Dirac delta $\delta(P^2-m^2)$ is included as a factor in the
momenta integral of (\ref{Fourier2}) {\it i.e.}, we obtain the
standard spacetime wave function satisfying the Klein-Gordon
equation (see (\ref{DiffReal-R6}) and (\ref{Klein-Gordon}))
 as well as space-time wavefunctions for arbitrary spins
 (see (3.21a--b)). In particular, if $s = \frac{1}{2}$, eqs. (\ref{Def-chi}),
(\ref{Def-chi inv}) lead to the standard Dirac equation in
ordinary Minkowski space, with gamma matrices written in the Weyl
realization.

\section{Final Remarks}\label{FinalRemarks}
\setcounter{equation}{0}

The aim of this paper is to exhibit the consequences of using the
geometric two-twistor framework for the simplest case of the
classical mechanics of free massive particles with spin. Our
starting point, not present in other discussions in the twistor
formalism (see {\it e.g.} \cite{Hu79}), is a rigorous derivation
of the particle action from the free two-twistor symplectic form
without introducing additional degrees of
 freedom\footnote{An example of the use of two-twistor geometry
  for the description of massive particles with spin
  which employs  additional degrees of freedom
  (the so-called index spinor) is provided
   by \cite{FZ}. This index spinor has been interpreted as a bosonic
   counterpart of the Grassmann components of the superstwistor
   for N=2 supersymmetry.}. In such a framework the mass, spin and electric
charge appear as free parameters of our model and are
 determined by additional non-geometric constraints.

One point in which our paper differs from the standard Penrose
framework is the definition of the spacetime coordinates.
Following the usual field-theoretic description of massive
particles with spin we introduce  commuting, composite spacetime
coordinates (see (\ref{XDef})). On the other hand, it is known
that the presence of spin leads naturally to non-commutative
spacetime  coordinates\footnote{We recall (see
(\ref{XX-PPConvPB})) that the non-commutativity of the spacetime
coordinates follows from the standard Penrose framework. The
quantization of a twistor-motivated particle model with
non-commuting spacetime coordinates ({\it i.e.} within the
standard Penrose definition of spacetime) will be considered in a
forthcoming paper \cite{Fe05}.} (see {\it e.g.} \cite{So70}), as
reflected by the non-vanishing of the PB among the spacetime
coordinates of the particles with spin; this is also the case when
spin is related to the presence of the supersymmetry
 and the superspace extension of classical mechanics
\cite{Ca76,AL83,Fe05} . We see therefore two options for
describing a two-twistor-inspired massive spinning particle
dynamics:

i) to use the standard Penrose approach with non-commutative
Minkowski coordinates satisfying the PB (\ref{XNonCommutative})
\cite{Fe05}. In such a case  to complete the description of the
quantized theory of particles with non-vanishing spin  one is led
to a field theoretic framework on non-commutative
spacetime\footnote{We point out that non-commutative field theory
has attracted considerable attention (see {\it e.g.}, \cite{DHR})
in the last decade, although such a non-commutativity does not
follow from the non-vanishing spin.}.

ii) to follow the approach used in this paper. In this case only a
partial quantization of the classical phase space degrees of
freedom is possible, and a complete quantization of all the Dirac
brackets, including those of the composite commuting spacetime
coordinates with the spin sector variables
($\eta_\alpha,\oeta_\dalpha, \sigma_\alpha,\osigma_\dalpha$),
requires further study.

Nevertheless, our present geometric framework assigns a dynamical
r\^ole to the additional twistor-motivated spinorial degrees of
freedom, where the momentum is expressed in terms of the
twistorial `constituent' variables. Although the use of auxiliary
fundamental spinor variables for the description of higher spin
theories is known (see {\it e.g.} \cite{Va99}), up to now most of
the twistor-based applications have been concerned with massless
fields with arbitrary helicity (see \cite{Ban00,Wi04,Ca04,Ca05}).
We would like to observe that the Cachazo-Svr\v{c}ek-Witten
twistor approach to maximally helicity violating vertices and tree
amplitudes in gauge theories \cite{Wi04,Ca04,Ca05} has recently
been extended to tree amplitudes including massive particles
\cite{Di04,Bad05}. Nevertheless, in such a framework the
description of massless four-momenta as composites of twistor
coordinates has not been extended to the four-momenta and spin for
massive spinning particles. We stress, however, that in the
approach presented here, which considers pairs of twistors, we
obtain a scheme that permits describing the four-momenta  as well
as the spin of massive particles as functions of twistorial
`constituent' variables. Also, our framework provides a
twistor-motivated approach to massive, free higher spin theory.

\textsl{Note added.} \\
The case of a massive spinless particle has recently been described
in terms of a single twistor by using a modified twistor-phase space
transform inspired by two-time physics techniques
\cite{BarsPicon05}.

\bigskip

\noindent
 {\it Acknowledgements}.
 This work has been supported by the Polish KBN grant 1 P03B 01828
 (A.F. and J.L.), the
 Spanish Ministerio de Educaci\'{o}n y Ciencia through the
 grant FIS2005-02761 and EU FEDER funds, the Generalitat Valenciana,
 the Poland-Spain scientific cooperation agreement and by the EU network
MRTN--CT--2004--005104. One of us (C.M.E.) acknowledges the
Spanish M.E.C. for his research grant. We also thank Sergey
Fedoruk for useful remarks.

\end{document}